\documentclass[acmtog,nonacm]{acmart}

\usepackage{booktabs} 
\usepackage{bm}
\usepackage{wrapfig}
\usepackage[
raggedright
]{sidecap}   
\sidecaptionvpos{figure}{t} 

\newcommand{\rev}[1]{#1}

\acmJournal{TOG}
\setcopyright{none}

\newcommand{\secref}[1]{Sec.~\ref{#1}}
\newcommand{\figref}[1]{Fig.~\ref{#1}}
\newcommand{\myeqref}[1]{Eq.~\ref{#1}}

\definecolor{darkgreen}{rgb}{0.0, 0.7, 0.0}
\newcommand{\new}[1]{{#1}}

\newcommand{\projectname}[0]{SkinCells}

\title{\projectname : Sparse Skinning using Voronoi Cells}

\author{Egor Larionov}
\orcid{0000-0002-9900-3150}
\email{elrnv@meta.com}

\author{Igor Santesteban}
\orcid{0000-0002-7276-083X}
\email{isantesteban@meta.com}

\author{Hsiao-yu Chen}
\orcid{0009-0007-5986-2093}
\email{hsiaoyu@meta.com}
\affiliation{%
\institution{Meta Reality Labs}
\country{USA}}

\author{Gene Wei-Chin Lin}
\orcid{0009-0006-5492-9941}
\email{genelin@meta.com}
\affiliation{%
 \institution{Meta Reality Labs}
 \city{Vancouver}
 \country{Canada}}

\author{Philipp Herholz}
\orcid{0000-0002-8389-792X}
\email{herholz@meta.com}

\author{Ryan Goldade}
\orcid{0009-0004-9336-8679}
\email{ryangoldade@meta.com}

\author{Ladislav Kavan}
\orcid{0000-0001-8549-0878}
\email{lkavan@meta.com}
\affiliation{%
 \institution{Meta Reality Labs}
 \city{Zurich}
 \country{Switzerland}}

\author{Doug Roble}
\orcid{0009-0004-3415-4283}
\email{droble@meta.com}

\author{Tuur Stuyck}
\orcid{0000-0003-1892-2137}
\email{tuur@meta.com}
\affiliation{%
 \institution{Meta Reality Labs}
 \city{Sausalito}
 \state{CA}
 \country{USA}}

\begin{abstract}

For decades, real-time skinning has been the cornerstone of character animation in visual effects and games. Despite its importance, the creation of animatable digital assets remains a labor-intensive manual process. Existing automated tools frequently struggle with intricate geometries, often necessitating significant manual refinement to reach production standards.
We present a robust, fully automated method for generating high-quality skinning weights from a standard mesh and skeleton in a canonical A- or T-pose. Unlike traditional approaches, our framework offers direct sparsity controls to limit bone influences per vertex -- a critical requirement for maintaining performance in large-scale mobile environments. Furthermore, we address the challenge of Level-of-Detail (LoD) management by optimizing weights within a continuous spatial volume rather than on discrete vertices. This allows a single optimization pass to be applied seamlessly across multiple asset resolutions and variations.
Central to our approach is a novel parameterized family of functions, we call \projectname{}. We demonstrate that our method consistently produces stable, high-quality results even in complex scenarios where standard biharmonic weight computations fail.

\end{abstract}



\begin{document}

\begin{teaserfigure}
  \includegraphics[width=\textwidth]{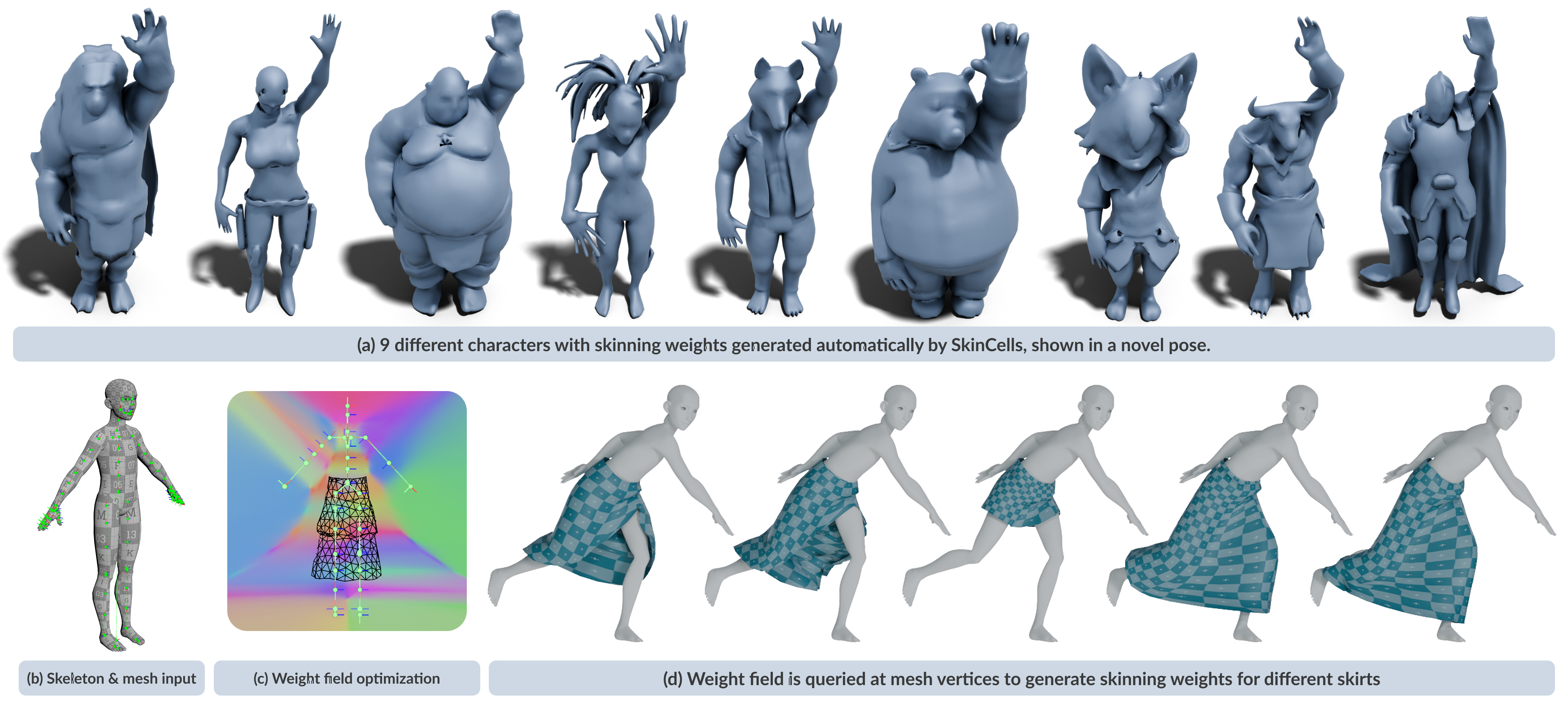}
  \caption{\projectname{}\ is a field-based framework for computing sparse skinning weights for characters and garments.
  By optimizing a weight field in canonical space (c), it binds meshes to a skeleton rig (b) while minimizing deformation artifacts.
  We demonstrate its effectiveness on nine characters using linear blend skinning (a) and five distinct skirt designs (d).  } 
  \label{fig:teaser}
\end{teaserfigure}

\maketitle

\section{Introduction} \label{sec:intro}
Character animation has long been a cornerstone of graphics research, with a myriad of techniques developed to efficiently animate complex shapes. A popular family of methods called \emph{skinning}~\cite{skinningcourse2014} is widely adopted in visual effects, animation, and gaming industries. These methods allow for the efficient manipulation of character models by binding them to skeletal structures; assigning a set of weights (scalar values that typically sum to 1) to each vertex, indicating which skeleton bones  the vertex should follow. This enables efficient real-time animation in interactive applications. Although automated skinning weight methods have been introduced, preparing these models for animation often still requires manual, labor-intensive processes. This is particularly true when dealing with complex characters that feature intricate details, accessories, and non-manifold geometries. In such cases, extensive manual refinement of automatically generated results is frequently necessary to achieve the desired level of quality.

Although physics-based simulations~\cite{wang2020adjustable, xu2016pose} and neural approaches \cite{yang2021s3} can offer a more realistic alternative, they are too resource-intensive for real-time use, especially on mobile devices that may need to animate and render many characters and assets simultaneously. In these scenarios, most of the limited compute budget is typically allocated for rendering since research shows that appearance modeling is key to achieving visual realism, e.g. for animated fabrics~\cite{aliaga2015}. 

The recent surge in generative AI models for automatically generating content at-scale highlights the growing need for dependable, high-quality, and automatic solutions that can efficiently produce animation-ready assets. While there is a large focus on mesh generation techniques~\cite{sarafianos2024garment3dgen, zhang2024clay, wei2024meshlrm, gao20243d}, these methods often omit generating accompanying rigs with skinning weights. To fully automate and democratize digital content generation, it is essential to address the full end-to-end process of rigged asset generation~\cite{liu2025riganything, Guo_2025_CVPR} and skinning weight computation.

In this work, we focus on automatic skinning weight computation, where the character mesh and a compatible skeleton have already been determined. Existing automated techniques for skinning weight computation, while innovative, often struggle with complex mesh topologies, details and LOD hierarchies found in production models. In response to these challenges, we introduce a novel approach for computing weights using a weight field representation designed to optimize skinning weights by leveraging randomly sampled character poses. This method reduces deformation artifacts and results in a weight field that is independent of mesh topology, directly providing weights for all LOD levels. Although one could resort to weight transfer across different resolutions, this requires additional steps and we demonstrate that transferring weights produces subpar results.

Similar to diffuse weights, a widely used technique for skinning 3D Gaussian Splatting (3DGS) \cite{kerbl3Dgaussians} representations of humans \cite{lin2022fite, li2024animatablegaussians}, our weight field representation is defined in ambient 3D space, allowing for immediate weight transfer across different models of the same character (e.g. to a high-fidelity model or a 3DGS representation).
Our weight representation is also compact, storing all parameters in single precision floats takes just 21 KB, whereas storing 4 weights per vertex along with corresponding weight indices can take up 11x more space in our experiments.

Our approach prioritizes robustness and high-quality deformations while maintaining sparse weight distributions, ensuring that the number of bone influences for each vertex can be constrained. This results in a highly efficient system capable of supporting numerous unique characters on devices with limited computational resources. \new{While prior work~\cite{Geodesic} has explored sparsity using falloffs to limit bone influences, it does not provide any numerical guarantees or controls on sparsity and optimizing for truly sparse weights while retaining smoothness has been acknowledged as an open problem \cite{skinningcourse2014}.}
We propose an optimization formulation that incorporates a novel set of goal functions to optimize skinning weights with the following desirable properties: smooth weights, ensuring a natural and continuous deformation of the mesh; sparsity for efficiency, reducing computational overhead by minimizing the number of non-zero weights, \new{essential for efficient asset creation, especially for performance-sensitive applications like mobile platforms \cite{DiamantSimantov2010}}; and an additional location objective, preventing trivial solutions where the mesh is not influenced by the skeleton, thereby maintaining a meaningful relationship between the two.

Our comprehensive evaluations reveal that our method strikes a balance between robustness and quality, generating high-quality weights while maintaining robustness. In contrast, comparative works often compromise on one aspect for the other, either prioritizing quality at the expense of robustness or vice versa. Our approach offers a streamlined solution for automatic skinning weight computation with sparsity enforcement and easy LOD weight generation, significantly reducing the time and effort required in the animation pipeline by offering a robust solution.

Compared to prior work, we provide \rev{the following contributions:}
\begin{itemize}
    \item Controllable enforcement for the limit on the number of bone influences per vertex, which we refer to as sparsity.
    \item A system that integrates with level-of-detail asset generation by optimizing spatially varying weights, applicable to all detail levels through a single optimization process.
    \item A novel optimizable weight field formulation inspired by Voronoi cell structures that is intrinsically diffuse, can enforce sparsity and works well for generating cloth and character skinning fields.
    \item A novel set of loss functions to produce optimal skinning weights. (i) A \emph{sparsity objective} to penalize weight distributions with more than a set number of nonzero weights per point. (ii) A \emph{smoothness objective} based on the DeltaMush~\cite{mancewicz2014deltamush} method. (iii) A spring-based \emph{location objective} ensuring that the mesh follows the skeleton.
\end{itemize}

\section{Related Work}

\paragraph{Skinning} Deforming a mesh through an underlying skeleton structure (i.e. skinning), is a longstanding, foundational approach for computer graphics animation \cite{magnenatthalmann1989lbs}. Recent research has focused on enhancing skinning accuracy through various techniques. \new{Considerable attention has been given to run-time improvements to skinning.
Dual quaternions~\cite{kavan2007dqs} improved rotation blending, and physics-based methods~\cite{projectiveSkinning} introduced physics priors into skinning. Other studies have explored the addition of second-order motions to standard skinning~\cite{ZhangCompDynamics2020, benchekroun2023FastComplemDynamics, rohmer2021velocity, kalyasundaram2022acceleration}. To simplify the process of editing skinning weights some works proposed to use splines~\cite{bang2018spline}, while we aim to automate this process altogether.  In some cases, optimal mesh deformations are known a priori, allowing algorithms~\cite{binhle2012} to find optimal rigging configurations to fit them.  Additional efforts have been made to improve efficiency for real-time high-quality deformations across multiple levels of detail (LOD)~\cite{zheng2024multi}, highlighting the importance of supporting LOD in animation. A current trend in machine learning involves optimizing weight fields by learning from physical models~\cite{modi2024simplicits} or directly from data~\cite{saito2021, liu2019neuroskinning, lin2022fite}. In either case, weight fields make a natural tool for building topology independent skinning with a natural support for LOD. \rev{Another approach is to predict vertex weights directly using neural models \cite{mosellamontoro2022skinningnet}, and some propose to predict skinning weights and hierarchical skeletons simultaneously \cite{xu2020rignet, song2025magic}.}
Another interesting line of work facilitates skin weight transfer~\cite{abdrashitov2023, cao2024skeleton}, which leverages template rigs to animate new characters or clothing.

\paragraph{Skinning Weights}
In this work, we focus on the automatic computation of skinning weights for a given skeleton rig with an emphasis on sparsity. A straightforward approach to computing skeleton-based weights assigns each mesh vertex a value inversely proportional to its distance from the skeleton. To limit the influence of distant bones, this value can be further attenuated with an exponent. This technique is commonly known as proximity-based weighting. While intuitive, this approach is blind to mesh topology and can produce artifacts where one limb can erroneously affect another if it is positioned close together (e.g. fingers on a hand) in a canonical pose.
To mitigate this, much effort has been dedicated to inject the mesh topology in the weight computation process. Diffusing bone influences to the surface mesh \cite{baran2007} helps resolve this artifact. Other works including \emph{bone glow} \cite{wareham2008}, \rev{\emph{quasi-harmonic} \cite{wang2021} and \emph{biharmonic} \cite{jacobson2011biharmonic, dodik2025robust} skinning} further improve the quality of generated weights. However, an important criteria for efficient skinning is maintaining sparsity of generated weights. While some methods \cite{landreneau2010, binhle2013} propose a post-process to compress a given set of skinning weights, \rev{others \cite{thiery2017araplbs} propose to enforce sparsity as a pre-process, which selects a fixed and potentially suboptimal configuration apriori. In contrast, our method has a built-in control for enforcing sparsity, which is optimized along with smoothness and locality penalties avoiding local minima especially within complex topologies. Prior work \cite{thiery2017araplbs, kavan2012} discovered the importance of penalizing deformation artifacts in sampled poses -- rather than enforcing intrinsic properties like weight smoothness. Our method builds on this idea giving the algorithm the opportunity to find elusive optimal weight distributions. Other works \cite{ma2024easyskinning} also leveraged the Voronoi nature of influence regions for building an interactive skinning tool.}
For a more comprehensive overview of prior work, see the course notes on skinning \cite{skinningcourse2014}.

 }

\paragraph{Animation} Physics-based simulation of soft tissues in characters~\cite{smith2018stable, neohookxpbd, kim2022dynamic} and clothing~\cite{BWCloth, stuyck2022cloth, choi2002stable} has been a long standing research field within computer graphics with many significant advances. It is an established solution for producing realistic body and garment behavior. An alternative to modeling pose-dependent deformations for a rigged body is to apply pose-dependent correctives. Although efficient, such a model needs to be manually set up or trained from captured data~\cite{allen2006learning, xu2020ghum}.

In recent years, the field has shown interest in learning-based solutions and skinning serves as the basis for many of these methods~\cite{bertiche2022ncs, santesteban2022snug, stuyck2024quaffure, stantesteban2019learningclothanim, halimi2022pattern}. These methods are conditioned on the body skeleton pose and predict deformations in a canonical pose, which gets posed to world space using skinning. This greatly facilitates learning as it provides a unified canonical space for easier learning.  Early work by \citet{stantesteban2019learningclothanim} proposed a data-driven approach to learning cloth dynamics, which was later extended to a self-supervised setting~\cite{santesteban2022snug, bertiche2022ncs} where cloth deformations are learned from physics supervision directly, without the need for simulated training data.

\paragraph{Voronoi fields} \citet{feng2023}  introduced a topology optimization algorithm that utilizes a differentiable and generalized Voronoi representation, enabling the evolution of cellular structures as continuous fields and extended this to accommodate anisotropic cells, free boundaries, and functionally-graded cellular structures. \citet{herholz2017voronoi} compute intrinsic Voronoi diagrams on surfaces using heat diffusion. As an application they compute skinning weights using Sibson coordinates. Based on a sparse set of point handles, weights for a specific vertex are computed by adding it to the diagram and analyzing how much area the new cell took from cells of the initial diagram.

\section{Method} \label{sec:method}
\newcommand{\x}{\mathbf{x}}
\newcommand{\s}{\mathbf{s}}
\newcommand{\p}{\mathbf{p}}
\newcommand{\y}{\mathbf{y}}
\newcommand{\T}{\mathbf{T}}
\newcommand{\F}{\mathbf{F}}
\newcommand{\R}{\mathbb{R}}
\newcommand{\Rot}{\mathbf{R}}
\newcommand{\diag}{\mathrm{diag}}
\newcommand{\LBS}{\mathrm{LBS}}
\newcommand{\w}{\mathbf{w}}
\newcommand{\X}{\mathbf{X}}
\newcommand{\Lap}{\mathbf{L}}
\newcommand{\B}{\mathcal{B}}
\newcommand{\V}{V}
Inspired by Voronoi diagrams, we similarly employ an implicit weight field function to automatically generate skinning weights. We begin by introducing the concepts of skinning (\secref{sec:skinning}) and Voronoi cells (\secref{sec:voronoi_diagrams}), laying the groundwork for the novel aspects of our method, which are presented in greater detail in the following sections. 

\subsection{Skinning} \label{sec:skinning}
Skinning is a fundamental technique in character animation used to deform a mesh according to an underlying skeletal structure, enabling realistic and controllable animation. 
One of the simplest and most widely adopted skinning techniques is Linear Blend Skinning (LBS) \cite{magnenatthalmann1989lbs, skinningcourse2014}. In LBS, each vertex of the mesh is moved by a weighted combination of bone transformations. Formally, the deformed position of a vertex $\x$ is computed as
\begin{align}
\LBS(\x; \w, \T) = \sum_{j = 1}^n \w_j\T_j\x, \quad \text{where} \quad \sum_j^n \w_j = 1, \label{eq:lbs}
\end{align}
where $n$ is the number of joints, $\w \in \R^n$ are the skinning weights and $\T_j \in \R^{4\times4}$ are the affine transforms of the joints in the current pose. This linear interpolation of transformations is computationally efficient and easy to implement, making LBS the \textit{de facto} standard in real-time applications. However, LBS suffers from several well-known artifacts, including volume loss near joints and unnatural collapsing or candy-wrapper effects during extreme bending. These limitations motivated more advanced skinning models, such as dual quaternion skinning \cite{kavan2007dqs} and pose-space deformation \cite{lewis2000psd} techniques, which aim to preserve volume and improve realism. However, these more complex techniques are applied at run-time, negatively impacting performance. 
We observe that, by leveraging the joint limits of the rig, we can optimize skinning weights as a pre-process using pose samples within these constraints to effectively reduce unnatural distortions and thus enhance overall deformation quality without incurring additional run-time costs. \new{Alternatively, poses can be sampled from motion capture or animated sequences but we found that randomly sampled poses produce comparable results while requiring fewer resources.}

\subsection{Voronoi Cells} \label{sec:voronoi_diagrams}

Voronoi diagrams are a construct in computational geometry that partition a given space based on proximity to a discrete set of points. Given a set $\{\p_1,\p_2,...,\p_n\}$
of distinct sites in a metric space (in our case $\R^3$), the Voronoi diagram divides the space into $n$ regions, such that each region $\V_D(\p_j) \subset \R^3$ consists of all points in the space that have $\p_j$ as their closest neighbour.
Formally, the Voronoi cell for site $\p_j$ is defined as $
\V_D(\p_j) = \{\x \in \R^3:\ d(\x, \p_j) \leq d(\x, \p_k) \forall k \neq j\}$,
where $d : \R^3 \to \R$ is a distance function.

Voronoi cells for constructing skinning weights follows the same motivations as traditional techniques; compute weights based on proximity to the nearest bone. This concept is also seen in diffuse skinning fields~\cite{lin2022fite, saito2021}, which resemble Voronoi cells with segment-shaped sites. By leveraging the link between heat diffusion and Voronoi diagrams, we can compute distance fields that mimic diffusion behavior~\cite{herholz2017voronoi}. 



\setlength{\intextsep}{5pt} 
\setlength{\columnsep}{5pt} 
\begin{figure}[t]
    \centering
    \includegraphics[width=0.99\linewidth]{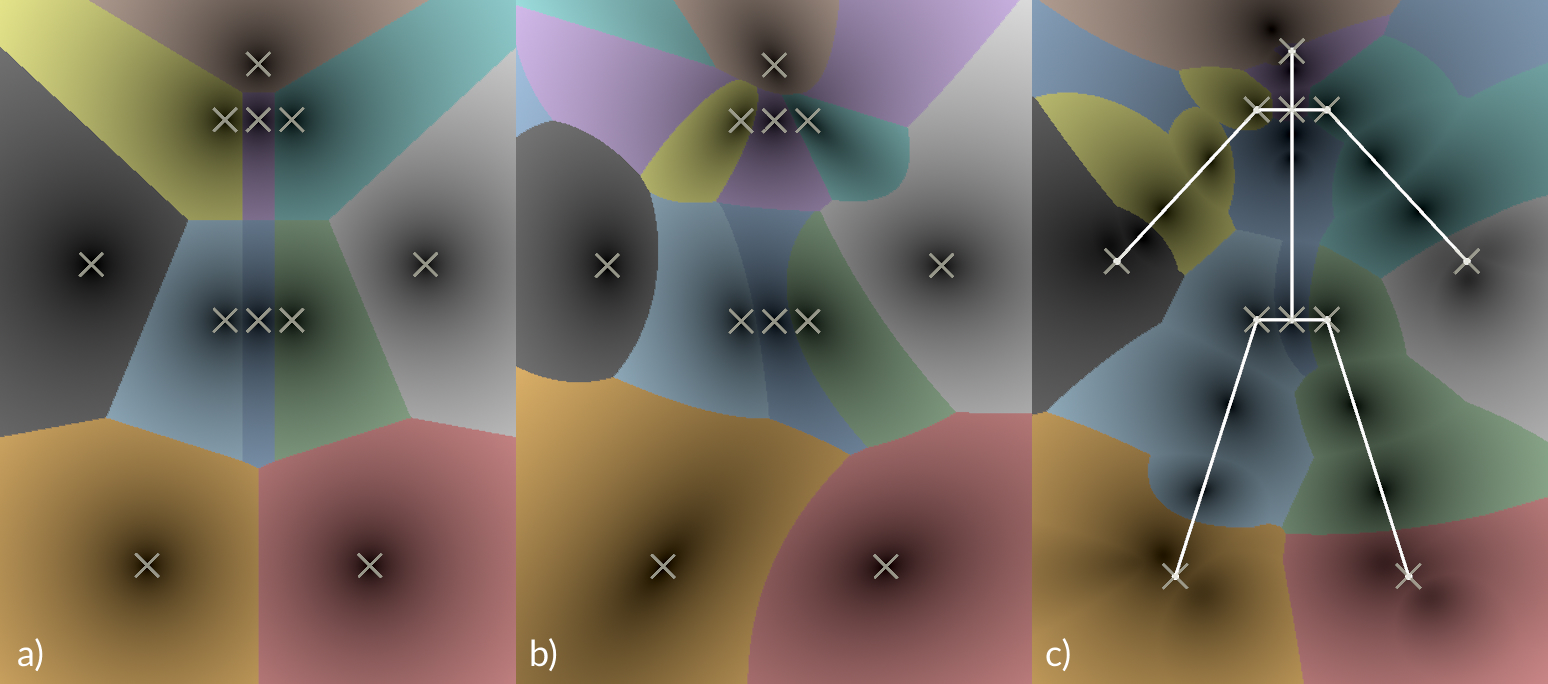}%
    \caption{A distance field (black at 0) for 11 regions (colored for clarity) marked with $\times$. (a) \rev{s}hows the field with uniformly shaped cells. (b) shows cells with randomly sampled $\s$ and $\Rot$ per site. In (c), each region is split into multiple sites sampled along the bones of the outlined skeleton.}%
    \label{fig:dist_field_shaped_mult}
\end{figure}


\subsection{Skin Cells} \label{sec:skin_cells}
Inspired by Voronoi cells, we propose a novel formulation called \emph{skin cells}. Voronoi diagrams create discrete boundaries of non-overlapping regions. Instead, skin cells introduces overlapping cells to control sparsity of the resulting field by adjusting the overlap of a predetermined number of cells. Each skin cell defines a local skinning weight field $w(\x; S_j)$  around each joint of the input skeleton, where $S_j$ is a set of parameters optimized per joint. We then combine these local fields to define the global skinning weight field as
\begin{align}
\label{eq:w}
\w(\x) = [\w_1(\x), ..., \w_n(\x)], && \text{ where } &&
\w_i(\x) = \frac{w(\x; S_i)}{\sum_j w(\x; S_j)}.
\end{align}
Since the field $\w(\x)$ is continuous and agnostic of the mesh topology, it can be easily applied to compute the skinning weights of multiple LoDs of the same mesh, or meshes that represent similar objects (e.g., reuse the same field to skin different types of skirts). While skinning weight fields have been extensively explored ~\cite{bhatnagar2020loopreg, saito2021, lin2022fite, santesteban2021collision}, our method provides a more compact parameterization where locality and sparsity are easily enforced. 

We first introduce our approach to model parametric distance fields by extending the Voronoi cell formulation, then we describe how to extract sparse skinning weights from the skin cell distances.

\paragraph{Distance function} We propose a \rev{Mahalanobis} distance metric from a point $\x$ to a site $\p$
\begin{align*}
    d(\x, \p; \F) := \sqrt{(\x-\p)^\top \F^\top \F (\x-\p)},
\end{align*}
where $\F\in\R^{3\times 3}$ is the deformation gradient of the site, i.e. $\F = \diag(\s)\Rot$ with $\s \in \R^3$ and $\Rot \in \mathrm{SO}(3)$. \rev{Note that $d$ is well defined since $\F^{\top}\F$ is symmetric positive definite (SPD).} This adaptation enables the metric to be scaled and rotated, allowing for a more nuanced and rig-aware weight distribution as shown in \figref{fig:dist_field_shaped_mult}.

\paragraph{Distance field}
To improve the ability to model complex distance fields, we allow the skin cells to have more than one site. We define the set of sites as $P_j=\{P_{j,1}, ..., P_{j,m}\}$ where $P_{j,k}$ are the site parameters including the site center $\p$, the scale $\s$, and the rotation $\Rot$, and $m$ is the user-specified number of sites. In our experiments, we find that $m=6$ is sufficient to strike a balance between performance and expressibility for our examples, although more complex meshes may benefit from a higher $m$. The distance to the skin cell $j$ is then defined as the distance to the closest site in $P_j$
\begin{align*}
    d(\x; P_j) &= \mathrm{min}_k\{h(d(\x; P_{j,k}), t_{j,k})\},
\end{align*}
\begin{wrapfigure}{r}{0.35\linewidth}
    \vspace{-2mm}
    \centering
    \includegraphics[width=\linewidth,trim={0 0 0 20},clip]{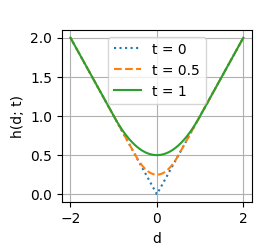}
    \label{fig:offset_huber}
\end{wrapfigure}
where $h$ is a function that modulates the contribution of each site to the distance field, controlled by an additional per-site parameter $t_{j,k}\in \R$. Inspired by the Huber loss~\cite{huber1992robust}, we define $h$ as shown in the inset with 
\begin{align*}
h(d; t) &= \left\{\begin{array}{ll}
\frac12 \left(\frac{|d|^2}{t} + t\right) & |d| < t \\
|d| & \text{otherwise.}
\end{array}\right.
\end{align*} 


\paragraph{Weight field} To extract valid skinning weights from the distance field, the distances must now be transformed into the range $[0, 1]$ with 0 being on the cell boundary and 1 at the site locations. We must also allow more than one nonzero weight per point in space, normalized to ensure weights add up to 1. To do this, first we define the pointwise sorting for these distances as
\begin{align*}
    D(\x) &= \mathrm{sort}[d(\mathbf{x}; P_j)]_{j=1}^n,
\end{align*}
where $D(\x)_1$, the first value, is the distance to the closest joint. 
Finally, we define the skinning weight field for each skin cell as
\begin{align}
w(\x; S_j) &= \left(\frac{\max\left(c_j,\, D(\mathbf{x})_l - d(\mathbf{x}; P_j)\right)}{d(\mathbf{x}; P_j)}\right)^{r_j} 
\label{eq:voronoi_field_fn}
\end{align}
where $S_j=\{P_j, c_j, r_j\}$ are the skin cell parameters including additional parameters $r_j$, $c_j$ that control the falloff and sparsity of the resulting field, and $l$ is the number of non-zero weights set by the user. Once the skin cell of each joint is defined, we compute the normalized skinning weights for a point $\x$ following \myeqref{eq:w}. \new{Note that \myeqref{eq:voronoi_field_fn} reduces to the standard proximity-based weight field when the numerator is set to a constant, which can be achieved for sufficiently large $c_j$.}

\begin{figure}[t]
    \centering
    \includegraphics[width=0.42\linewidth]{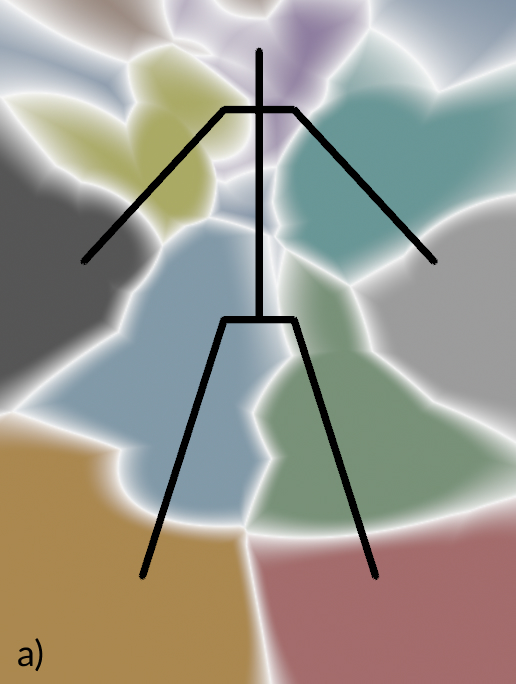}%
    \includegraphics[width=0.42\linewidth]{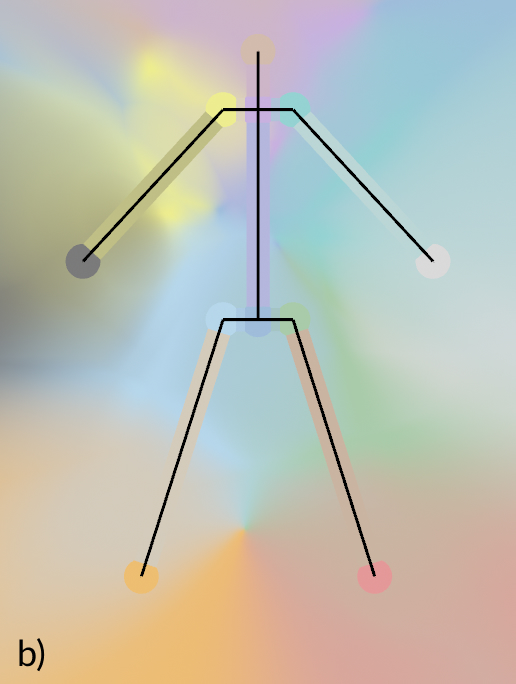}%
    \caption{(a) shows the unnormalized weight field $\w$ for $l=1$ where each region $j$ is uniquely colored and white indicates a zero value. (b) plots the normalized Voronoi field with $l=3$ indicating that there are at most 3 nonzero weights for each point $\x$. The outline of the source skeleton is shown in black for clarity.}%
    \label{fig:final_initialized_field}
\end{figure}

\paragraph{Sparsity}
Examining the fractional term in \myeqref{eq:voronoi_field_fn} we can see that for $c_j = 0$, we have $\|\w(\x)\|_0 \leq l$ for all $\x$ where $\w$ is well defined, where $\|\cdot\|_0$ is the $\ell_0$-``norm'' counting nonzeros of a vector. This guarantees that for any set of parameters, setting $c_j = 0$ for all $j$ produces a guaranteed sparse field with $\w(\x)$ having at most $l$ nonzeros for any $\x$.  The purpose of $c_j$ as an optimizable parameter is to relax the sparsity constraint temporarily to enable non-zero objective gradients in distant points during optimization. \figref{fig:final_initialized_field}a shows the unnormalized value of $\w$ for $l=1$, where the boundary between the cells is clearly visible. 
\figref{fig:final_initialized_field}b plots $\w$ for $l=3$ initialized using a toy skeleton and the scheme described in the next section.

\subsection{Initialization} \label{sec:init}


To improve parameter optimization convergence, we must first distribute the sites for each skin cell. The skin cell sites are initialized by sampling along the hierarchical skeleton rig, composed of 3D points representing joints and connecting segments representing bones. Each joint corresponds to a single skin cell with $m$ sites. For each joint with exactly one child, we sample $m$ points along the bone. For joints with multiple children we randomly sample the \emph{convex hull} containing all the immediate child joints. Finally, the sites associated to leaf joints (like ends of fingers or feet) are sampled in a small $0.5$ mm radius around the joint.
The remaining parameters are sampled from a uniform distribution. \rev{Rotations $\Rot$ and scales $\s$ can be sampled separately from a uniform distribution, or alternatively, the lower triangular entries of $\F$ can be randomly sampled directly spanning the space of SPD matrices $\F^{\top}\F$.} Parameters $c$, $t$, $r$, and are uniformly sampled from $\text{exp}(\text{Uniform}(-0.05,0.05))$, where. In practice, to enforce strict positivity, we apply the exponential on the optimized parameters before passing them to \myeqref{eq:w}.

\subsection{Loss Formulation} \label{sec:loss}
An optimal set of weights for a given mesh and skeleton should minimize the amount of mesh deformation caused by skeleton poses, while preserving the overall \emph{intent} of the pose. In other words, the mesh should follow the skeleton while preserving a desired level of smoothness. Furthermore, we aim to produce skinning weights of no more than $l$-many bone influences per vertex. This motivates the three losses we propose to minimize (see \secref{sec:opt}) over skin cell parameters $S_j$ to achieve the best possible skinning weight distribution.

\paragraph{Smoothness loss} Inspired by traditional graphics techniques, we reformulate a popular method for smoothing deformations, called DeltaMush \cite{mancewicz2014deltamush}, into a form that can easily fit our optimization problem. The method was originally designed as a run-time filter to smooth away artifacts introduced by skinning while retaining surface details present in the rest geometry. This is done by adding the delta produced by smoothing in rest configuration, to the result of smoothing the deformed configuration. This is performed in the coordinate frame of the surface to preserve as much detail as possible. The method relies on Laplacian smoothing, an iterative procedure of repeated applications of the discrete linear Laplacian $\Lap : \R^{3V} \to \R^{3V}$ on $V$ mesh vertex positions. 

We represent a mesh configuration using a matrix of stacked vertex positions. Specifically, we use $\bar{\X} \in \R^{3V}$ to denote the rest configuration of the mesh and $\X = \LBS(\bar{\X}; \w, \T)$ the deformed configuration after skinning, as described in \myeqref{eq:lbs}. A single iteration of the Laplacian operator $\Lap$ is sufficient for our purposes, so the smoothed configuration can be defined as $\X' = \X + \lambda_{\text{DM}}\Lap\X$, where $\lambda_{\text{DM}}$ is the smoothing coefficient that is later used to scale the resulting loss component. For brevity, we denote the change of basis rotation that maps vectors from $\bar{\X}$ to the deformed configuration $\X$ as $\B$. DeltaMush can then be written as $\X_D = \B(\bar{\X} - \bar{\X}') + \X'$, which reduces to
$\X_D = \X + \lambda_{\text{DM}}(\Lap(\X) - \B \Lap(\bar{\X}))$.
This expression indicates that the displacement produced by DeltaMush can be defined as $\lambda_{\text{DM}}(\Lap(\X) - \B \Lap(\bar{\X}))$. Instead of applying this displacement at run-time, we use it to formulate the following loss that encourages smooth deformations while preserving surface details
\newcommand{\LDM}{\mathcal{L}_{\text{DM}}}
\begin{align}
    \LDM = \|\Lap(\X) - \B \Lap(\bar{\X})\|_2^2. \label{eq:ldm}
\end{align}
\newcommand{\ds}{\mathbf{d}_{s}}


\paragraph{Location loss}
The minimizer of just \myeqref{eq:ldm} is the rest configuration $\X$ which does not follow the intended skeleton animation. This means that minimizing a smoothness loss alone will not produce a useful weight distribution. 
This phenomenon has been previously addressed by adding an orthogonality loss on the weight distribution \cite{modi2024simplicits}. However, in our case it is useful to preserve the weight field correspondences to bones, which makes the result editable and controllable in downstream applications. To ensure that the mesh follows the skeleton, we attach springs from each vertex on the mesh to the closest point on the skeleton. We prune any springs not orthogonal to the connecting bone and that intersect the mesh in any way. Let $\ds : \R^{3V} \to \R^M$ denote the distances of all $M$ valid springs for a given mesh configuration. The resulting loss can then be written as
\newcommand{\Lloc}{\mathcal{L}_{\text{loc}}}
\begin{align}
    \Lloc = \left\|\frac{\ds(\X) - \ds(\bar{\X})}{\ds(\bar{\X}) + \rev{10^{-2}}}\right\|_2^2, \label{eq:lloc}
\end{align}
where the denominator (computed coordinate-wise) ensures that far away vertices are not constrained as much as vertices close to the skeleton as one would expect on a real body. \rev{The $10^{-2}$ stabilizer appearing in the denominator assumes units to be in cm.}

\paragraph{Sparsity loss}
\newcommand{\topk}{\mathrm{Top}}
Finally, we encourage sparsity using a loss that indirectly penalizes artifacts caused by clamping the lowest $n - l$ weights in $\w$ to 0. Formally, if we define $\topk[l] : \R^n \to \R^n$ to set the lowest $n - l$ weights to zero such that $\|\topk[l](\w)\|_0 \leq l$, then our sparsity loss can be written as
\newcommand{\Lsp}{\mathcal{L}_{\text{sp}}}
\begin{align}
    \Lsp = \left\|\LBS(\bar{\X};\w) - \LBS(\bar{\X};\topk[l](\w))\right\|_2^2. \label{eq:lsp}
\end{align}
With positive $c_j$ the field function in \myeqref{eq:voronoi_field_fn} allows joints to have some small influence on far away vertices. This produces a non-zero gradient, even for bones that are farther away giving the optimization a better chance at recovering from a poor initialization.

The final loss is a weighted sum of these three losses
\newcommand{\Loss}{\mathcal{L}}
\begin{align}
    \Loss = \lambda_{\text{DM}}\LDM + \lambda_{\text{loc}}\Lloc + \lambda_{\text{sp}}\Lsp, \label{eq:l}
\end{align}
where we choose $\lambda_{\text{DM}} = \lambda_{\text{sp}} = 1$, and $\lambda_{\text{loc}} = 6000$ for all our experiments (unless otherwise stated). \rev{We found that a large range of parameters generates reasonable results. For details, refer to \secref{sec:results} under \emph{Smoothness and control trade-off}. When skinning fine meshes, it may be necessary to prioritize sparsity to avoid tearing artifacts. For instance we used $\lambda_{\text{sp}} = 1000$ for the example in \figref{fig:lod}. Note that while the space of parameters is scale independent since we sample distances on the mesh scaled to a unit box, the loss \myeqref{eq:l} remains scale sensitive.}


\subsection{Optimization} \label{sec:opt}
\begin{figure}
    \centering
    \includegraphics[width=0.8\linewidth]{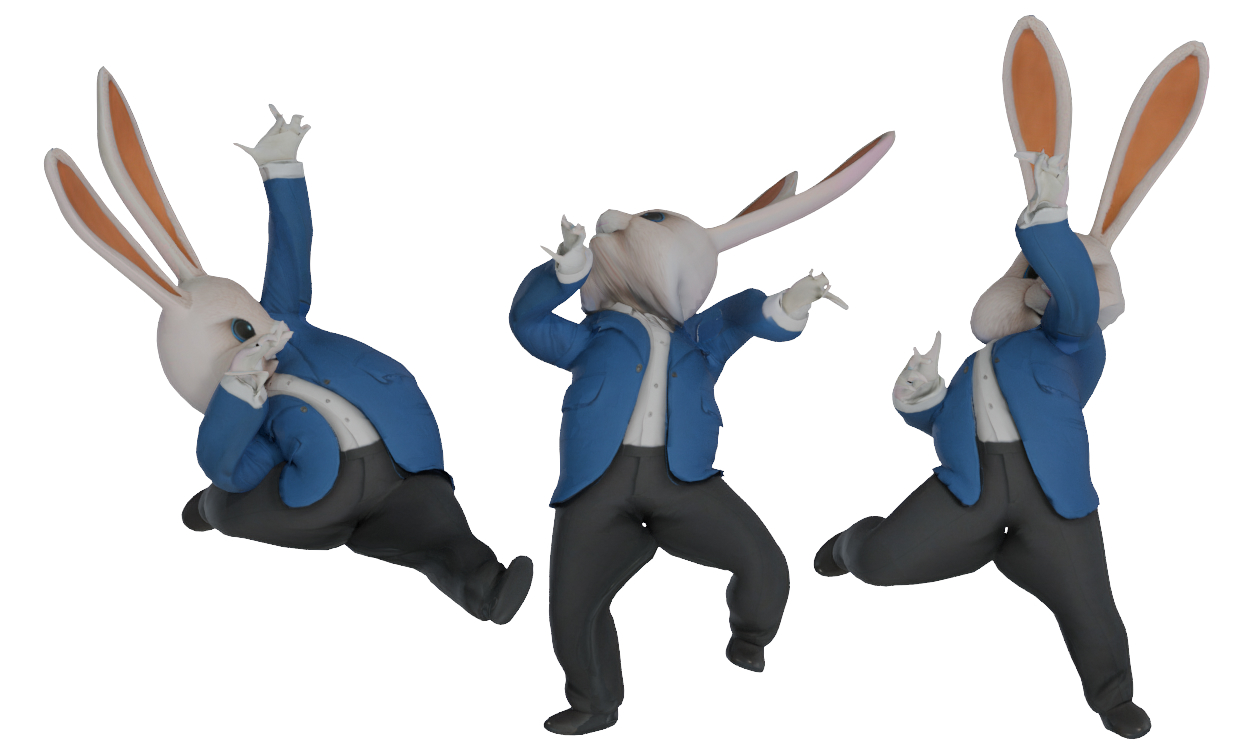}%
    \caption{Three examples of uniformly sampled poses used for further optimizing $\w$ by updating $S$ to minimize $\Loss$.}%
    \label{fig:random_poses}
\end{figure}

Once the weight field $\w$ is initialized as described in \secref{sec:init}, we optimize the skin cell parameters $S={S_1, ..., S_n}$ by evaluating the loss $\Loss$ at randomly sampled poses.
Taking a manually defined range of motion for each joint on a hierarchical rig, we uniformly sample joint angles $\bm{\theta}$
to produce random (and usually non-physical) poses as shown in \figref{fig:random_poses}. Alternatively, one could sample from animated sequences when available.
\newcommand{\J}{\mathbf{J}}
We define the joint transform $\J : \R^{3n} \to \R^{n \times 4 \times 4}$, which transforms local joint angles to world space transforms $\T$, which gives the final optimization problem:
\begin{align}
    \arg\min_{S} \Loss(\w(S), \J(\bm{\theta})). \label{eq:opt}
\end{align}

\section{Results} \label{sec:results}
We implement the optimization problem defined by~\myeqref{eq:opt} in PyTorch using the Adam optimizer. We sample 20 poses per batch and set the learning rate to $10^{-3}$.
For all experiments we set the maximum number of bone influences $l=4$, however larger values can be used to obtain smoother results allowing more joints to affect the motion of each vertex.

\begin{figure}
    \centering
    \includegraphics[width=0.7\linewidth]{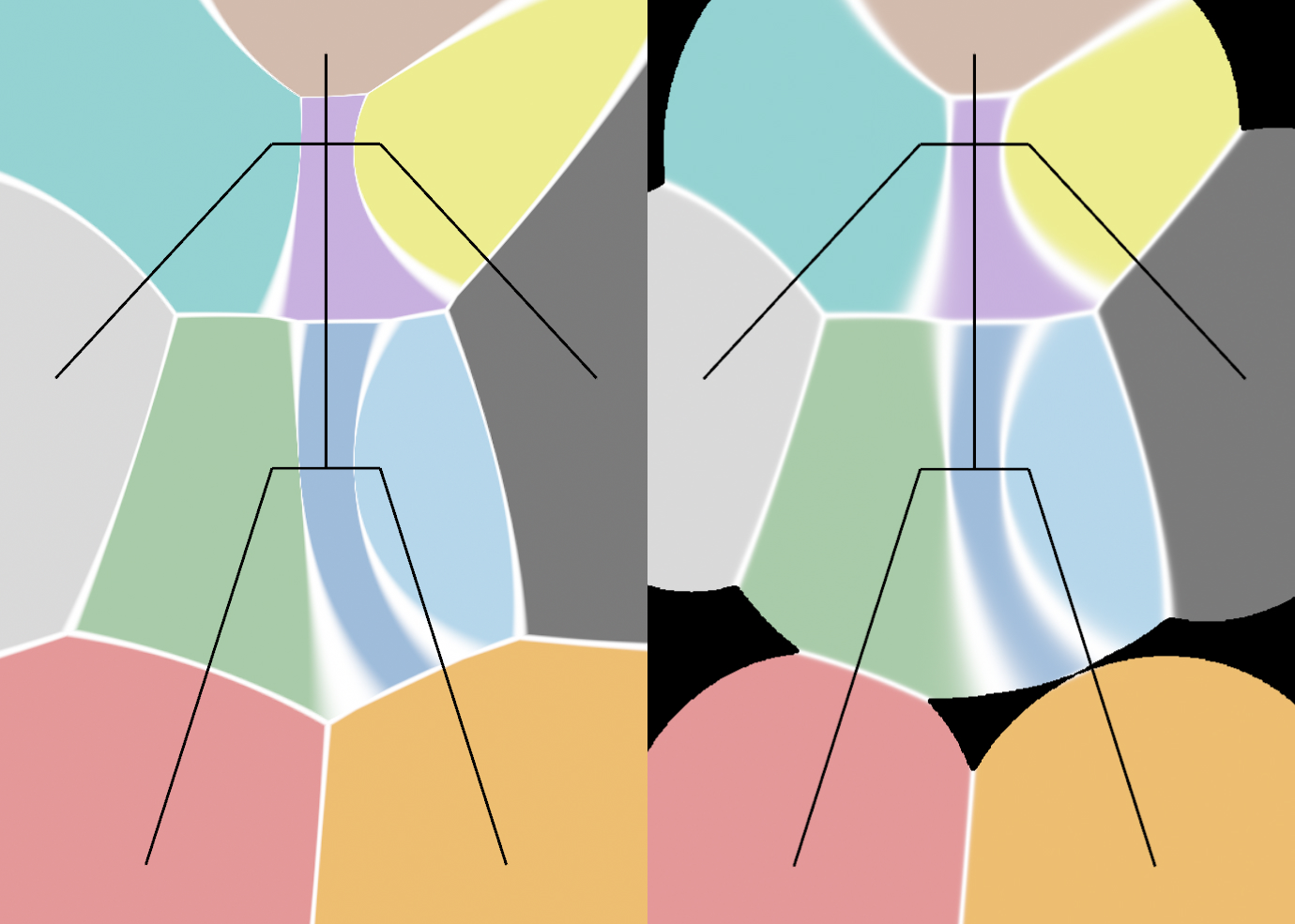}
    \caption{The Voronoi field produced by our method (left) does not suffer from the same numerical instability resulting from softmax used by prior methods (right) where black \rev{represents} NaN regions \cite{feng2023}.}%
    \label{fig:stability}
\end{figure}

\paragraph{Voronoi field stability}
Prior work has employed smooth Voronoi fields~\cite{feng2023} to represent differentiable Voronoi diagrams for representing cellular structures by applying softmax on the distance field, thereby generating a normalized and smooth field with a controllable falloff. However, this method runs into numerical instabilities even for moderately sized exponents, which is especially noticeable as distance values increase. In \figref{fig:stability}, we show how our field function compares to the state-of-the-art differentiable model proposed by \citet{feng2023} for sites seeded using a 2D toy skeleton. The rational expression used in softmax divides small floating point numbers for larger distances. This is exacerbated when using a larger exponent to achieve a sharper falloff, which results in numerical instability shown by the black regions in the figure.

\begin{figure}
    \centering
    \includegraphics[width=0.99\linewidth]{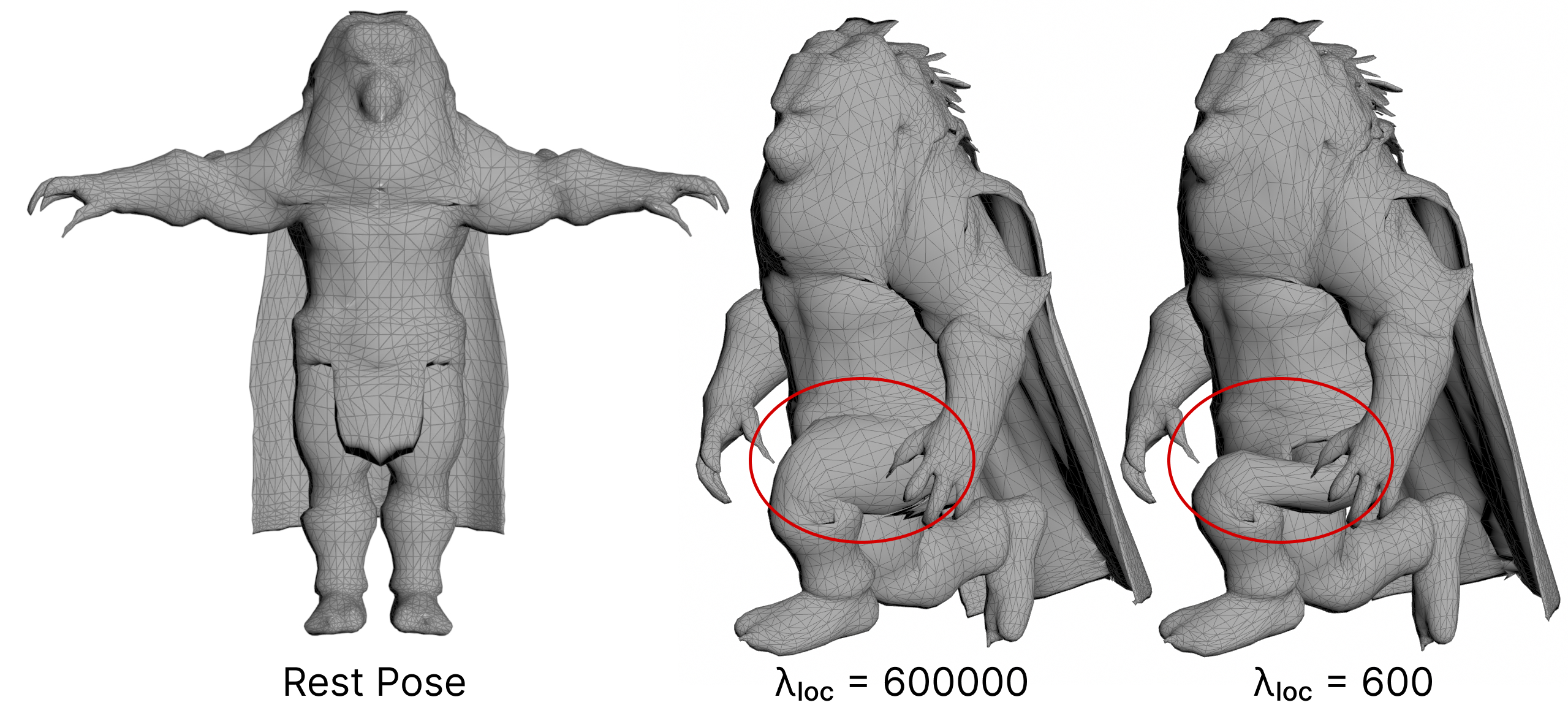}
    \caption{The parameter $\lambda_{\text{loc}}$ is increased (middle) on the lion character (model in rest pose on the left) to produce deformations more faithful to the underlying skeleton. Decreasing $\lambda_{\text{loc}}$ (right) produces a smoother fold where the leg bends, albeit with more volume loss.}%
    \label{fig:parameter_tuning}
\end{figure}

\begin{figure*}[t]
    \centering
    \includegraphics[width=\textwidth, trim={0 3 0 0}, clip]{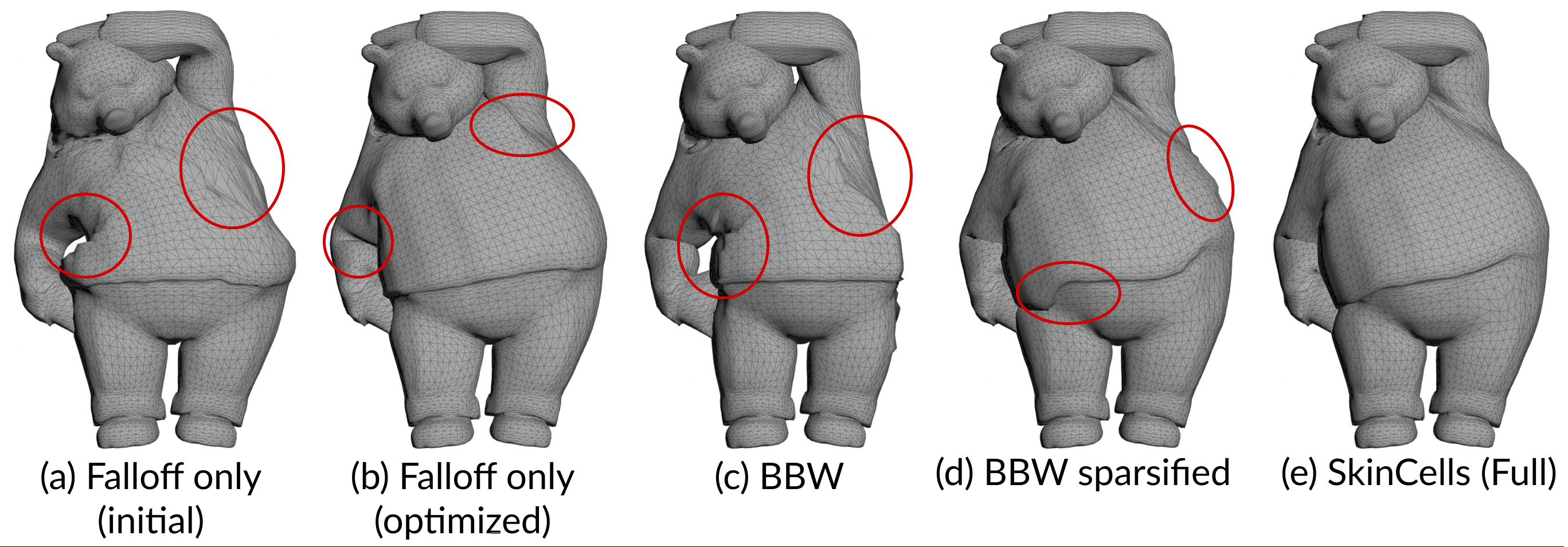}
    \caption{\new{We compare our method against others for optimizing skinning weights. In all cases bone influences are capped at 4 per vertex. The left two examples show the result of optimizing just the  falloff coefficient $r_j$, while keeping all other parameters fixed. The figure marked with (initial) corresponds to the traditional proximity-based approach for computing skinning weights, showing significant artifacts. By optimizing just the falloff with our approach, we can already eliminate most sparsity artifacts. We also compare against the bounded biharmonic weights (BBW) method (c), which also shows artifacts since it lacks any sparsity controls. If we further optimize the BBW weights on each vertex to minimize the loss in~\myeqref{eq:l} as shown in (d), we can eliminate some of the artifacts, although not all due to problematic local minima. Finally, the full SkinCells method produces a smooth result avoiding local minima and much more quickly than by optimizing falloff alone.}}%
    \label{fig:compare_bbw_falloff}
\end{figure*}

\paragraph{Smoothness and control trade-off}
Our formulation has a unique property that users can adjust the trade-off between between smoothness of the resulting skinning and control (how closely the mesh follows the underlying rig) by adjusting the $\lambda_{\text{loc}}$ parameter. In \figref{fig:parameter_tuning} we show how increasing the $\lambda_{\text{loc}}$ parameter results in deformations that more faithfully follow the underlying skeleton (e.g. in the leg of the character (middle)), while decreasing the parameter results in smoother deformation (e.g. the leg is smoother along the bends but looses volume (right), compared to the rest shape (left)).

\paragraph{Timing results}
Optimizations were run for 1500 steps with 1024 \rev{randomly sampled} poses batched at 16 poses per iteration. \rev{On an RTX A6000 GPU, in all tried examples the relative change in deformation became minimal within 1 minute, indicating convergence.}

\subsection{Comparisons}
We compare our method to two main approaches, a proximity-based weight mapping and bounded biharmonic weights (BBW)~\cite{jacobson2011biharmonic}. We demonstrate that while the proximity-based method is robust, results are not always sufficient and can lead to artifacts.
On the other hand, BBW produce high quality results but lack robustness (due to the dependency on a tetrahedralization step) and sparsity controls. We show that our method provides a robust solution to obtaining high quality weights.

\begin{figure}
    \centering
    \includegraphics[width=0.95\linewidth]{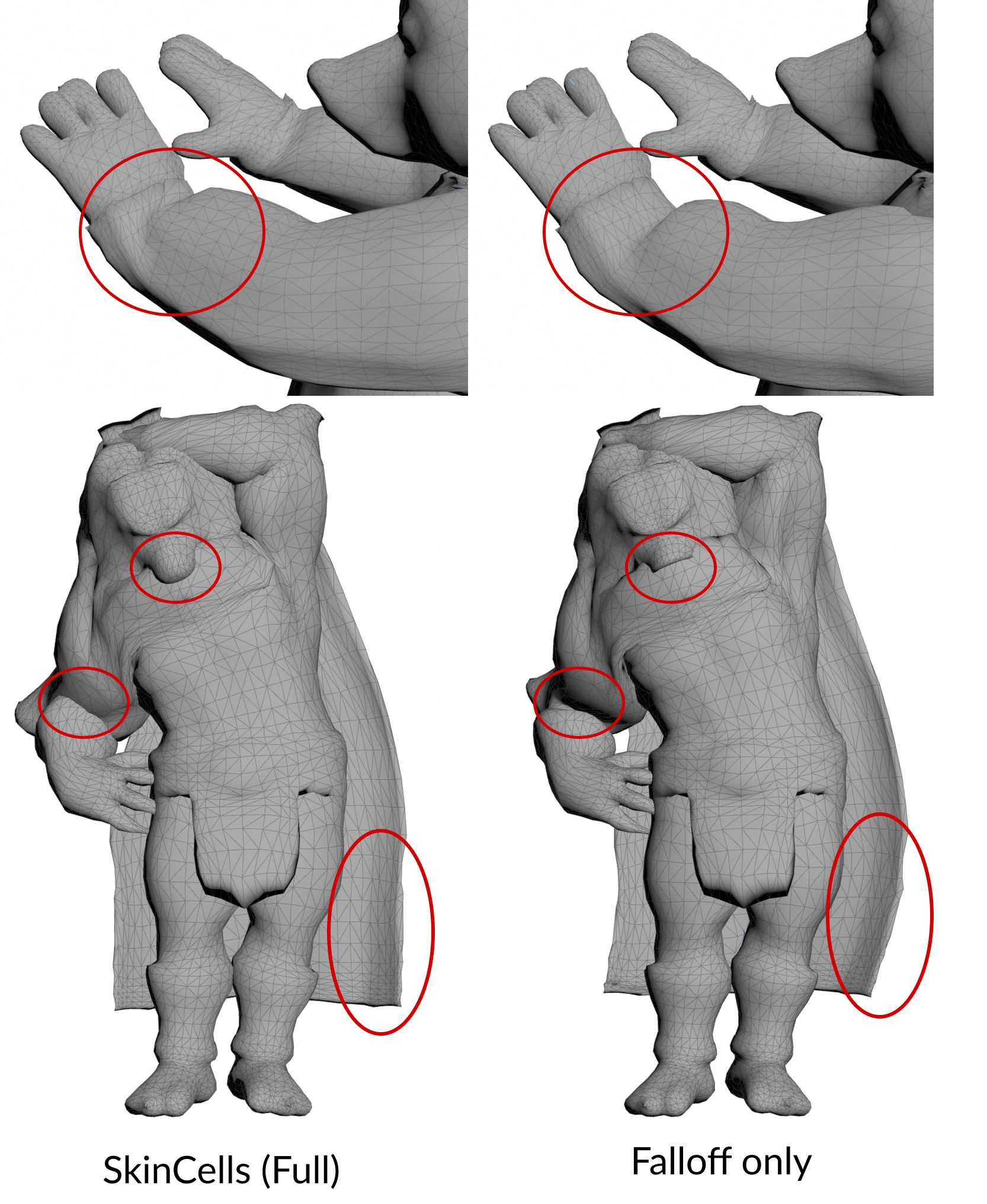}
    \caption{\new{Optimizing only the falloff can produce smooth results but without optimizing the location and shape of the cells, the field may exhibit various deformation artifacts if the underlying skeleton is not sufficiently expressive. The additional parameters in SkinCells can compensate for limited rigging as shown on the right.}}%
    \label{fig:compare_falloff}
\end{figure}

\begin{figure}
    \centering
    \includegraphics[width=\linewidth]{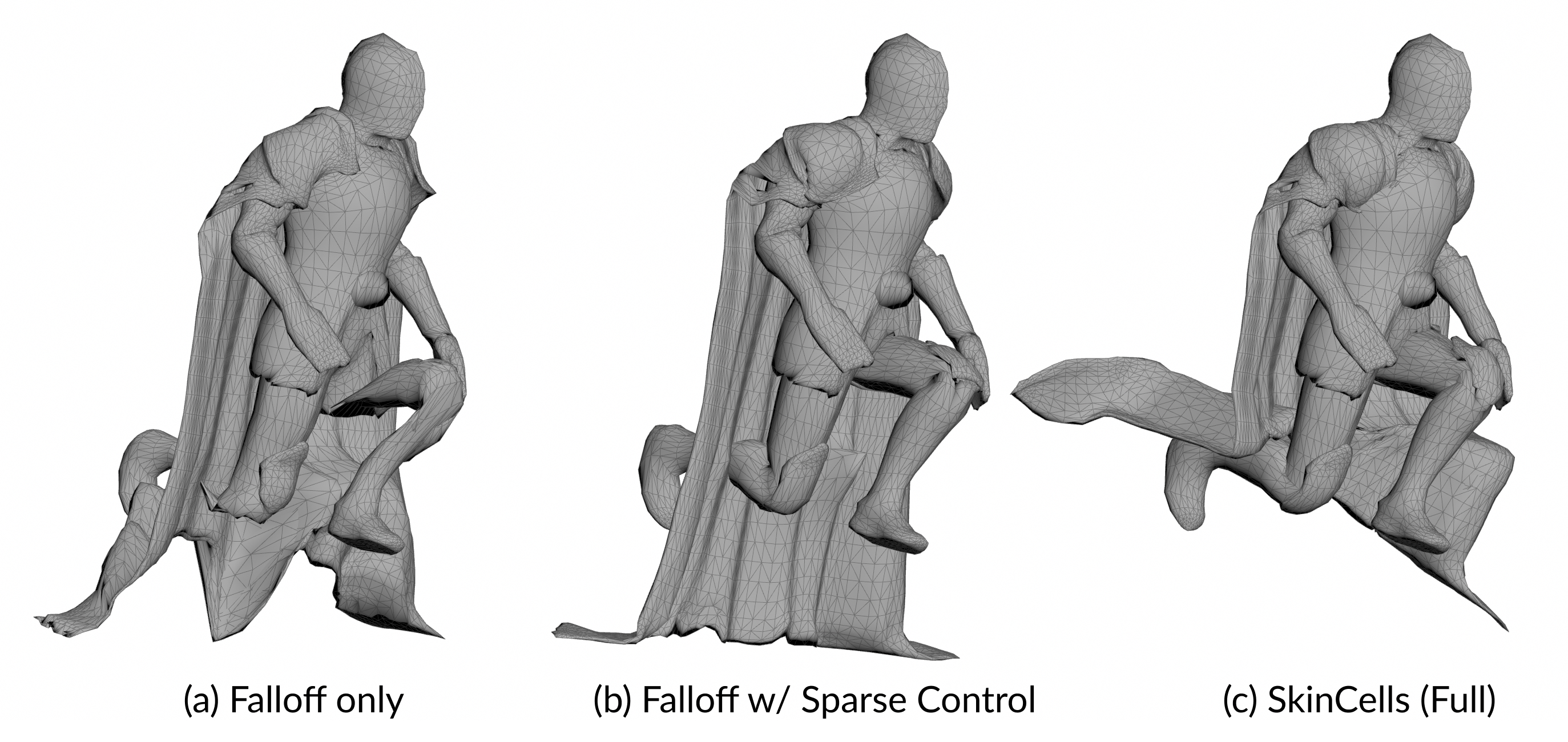}
    \caption{\new{The knight character is a particularly difficult example for skinning since the cape has no dedicated rig. As a result, without additional controls the optimization fails to find a suitable set of falloff coefficients to minimize the loss (a). Just adding the sparsity control in the numerator of  \myeqref{eq:voronoi_field_fn} can already help exclude the cape from being influenced by the legs (b). The full SkinCells model retains this ability maintaining good quality deformation (c).}}%
    \label{fig:knight}
\end{figure}

\begin{figure}
    \centering
    \includegraphics[width=\linewidth]{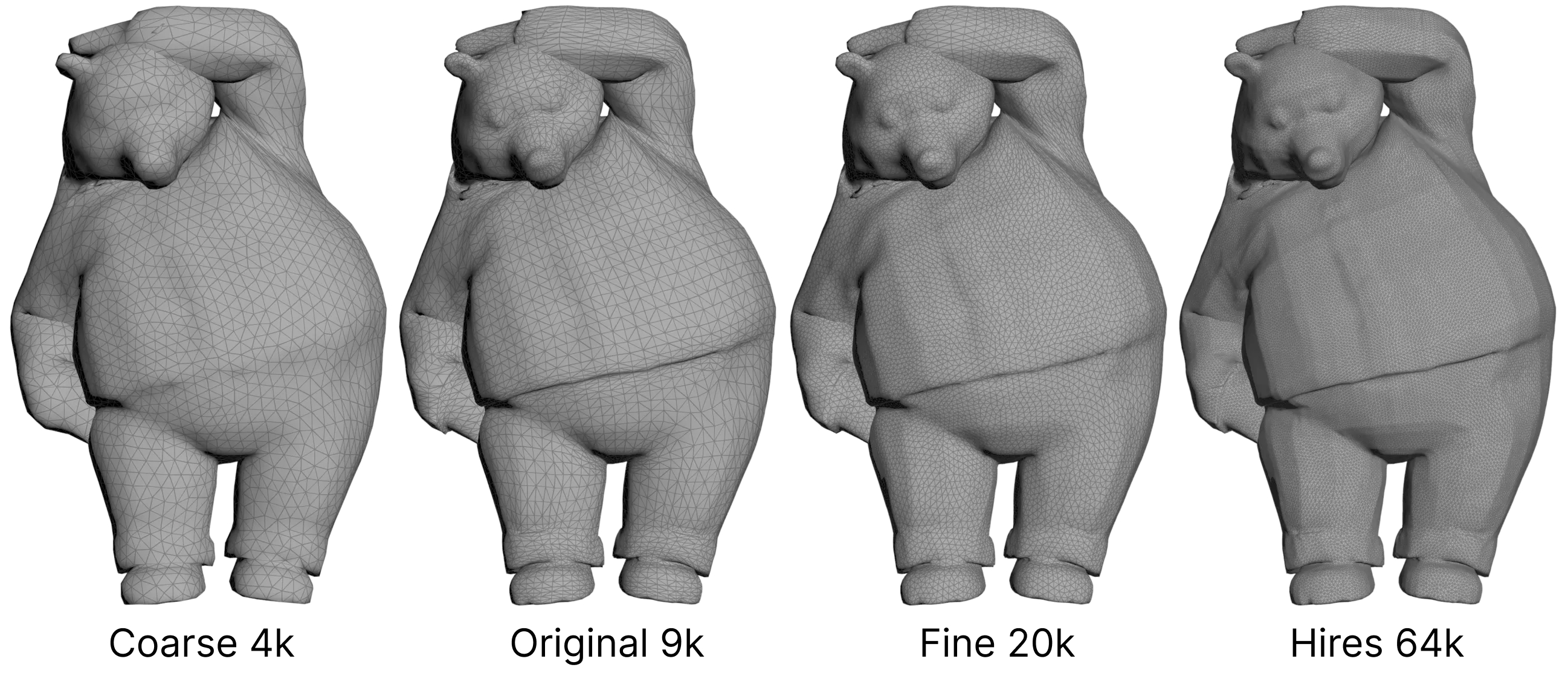}
    \caption{\rev{The bear model in a stretching pose at 4 different resolutions showing one optimization result applied at various levels of detail. The optimization used the ``Original'' model, and the resulting weight field is evaluated at the vertices of the 4 different meshes in canonical pose to generate the final vertex weights. The labels indicate the approximate number of vertices in each mesh.}}%
    \label{fig:lod}
\end{figure}

\begin{figure*}
    \centering
    \includegraphics[width=\linewidth]{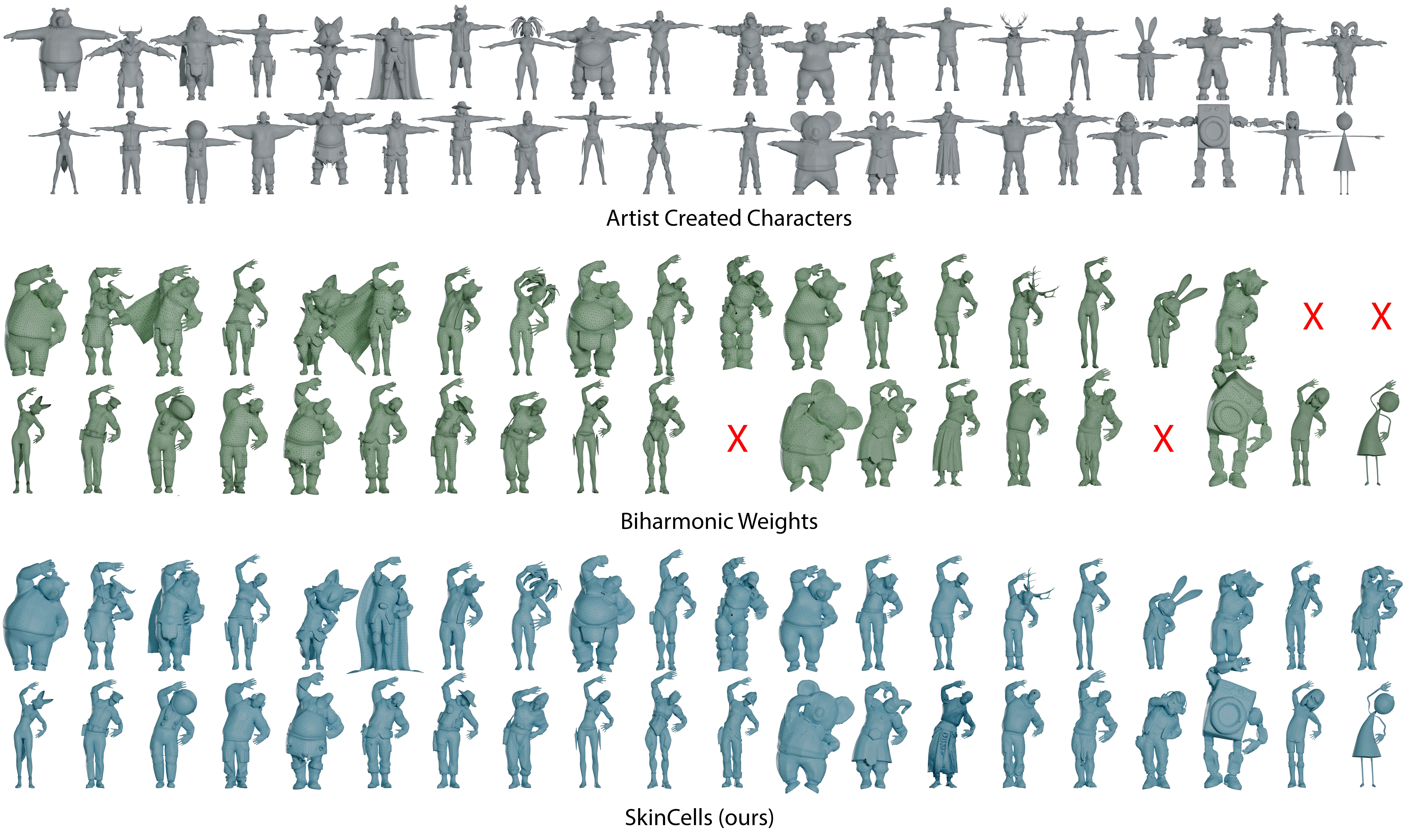}
    \caption{40 artist created characters (top) are skinned with 2 different methods. 36 out of 40 characters are skinned using bounded Biharmonic weights \cite{jacobson2011biharmonic} (middle). 3 models fail to tetrahedralize and for 1 the biharmonic solve failed for all tried solve parameters. In contrast, all 40 characters were processed automatically and successfully with our method (bottom). }
    \label{fig:full_comp}
\end{figure*}

\paragraph{Character skinning}
We apply our method to a set of 40 artist generated character models with associated skeleton rigs, and compare against BBW \cite{jacobson2011biharmonic} as implemented in Houdini 20.5 \cite{sidefx2025}. Each of the models was reposed in a T-pose, with legs spread. This can be done with any suitable method to smoothly separate the limbs into a specific pose like Laplacian mesh editing \cite{sorkine2004lap} or as-rigid-as-possible deformation \cite{sorkine2007arap}. For convenience, we used a different heuristic of deforming the character mesh using temporarily-transferred skinning weights from a fitted pre-skinned template. Although not strictly required, this ensures minimal self-intersections between the arms and torso and between the legs, producing a tetrahedralization with the fewest pruned intersecting triangles (required for Houdini's tetrahedralizer), giving the best quality mesh and thus optimal biharmonic weights. This pose is also ideal for spatial weight methods \cite{lin2022fite, saito2021, li2024animatablegaussians} like our approach, since it avoids significant weight contributions between adjacent limbs.
In 
\figref{fig:full_comp} we show the same pose for all skinned characters, and indicate where the method failed due to poor mesh quality (failed tetrahedralization). Notably, biharmonic weight computation failed on several 
characters, whereas our approach successfully computed weights for all. Furthermore, our method is able to produce good quality deformation while enforcing sparsity, which is especially noticeable for larger models with vertices far from the skeleton. For most other characters where biharmonic weight computation was successful and produced reasonably sparse fields, we observe comparable weight quality to our method.

\new{
\paragraph{Comparison case study}
In \figref{fig:compare_bbw_falloff}, we compare different approaches for computing weights on a single bear character.
Recall that \myeqref{eq:voronoi_field_fn} can be simplified to $w(\x; S_j) = 1 / d(\mathbf{x}; P_j)^{r_j}$, which reduces SkinCells to a proximity-based method if all other parameters a fixed, $t_{j,k}$ set to 0, $\F$ set to identity reducing to a Euclidean distance, and cell sites placed along skeleton bones. Setting the falloff $r_j=4$ results in the deformation shown in \figref{fig:compare_bbw_falloff}a, which exhibits significant artifacts. We can optimize just the falloff by minimizing the loss \myeqref{eq:l} over $r_j$, which produces the smooth result in \figref{fig:compare_bbw_falloff}b. While this case produces a compelling result, the deformation can exhibit volume loss for complex geometries as further investigated in the \emph{Ablations} section below. Out-of-the-box BBW algorithm produces tearing artifacts when weights are clamped to the largest 4 as shown in \figref{fig:compare_bbw_falloff}c. However, this can be mitigated using compression (or sparsification) algorithms \cite{landreneau2010, binhle2013} by optimizing per-vertex weights to be sparse. In \figref{fig:compare_bbw_falloff}d, we used the same loss from \myeqref{eq:l}, to produce a smooth result, however regardless of the optimization technique, this approach struggles to resolve local minima when the weights are not initially sufficiently sparse. The full SkinCells approach in \figref{fig:compare_bbw_falloff}e generates a good quality result satisfying sparsity and smoothness, and it converges approximately 5x faster thanks to the abundance of tunable parameters available to SkinCells \rev{beyond} just falloff.

\paragraph{Ablations}
We further investigate the limitations of reducing SkinCells to optimize only the falloff as described in the previous paragraph. In the top row of \figref{fig:compare_falloff} we highlight that the falloff-only  optimization fails to capture the wrist bend since weights behave isometrically and so are not able to better fit the underlying geometry to minimize deformation artifacts. In the bottom row of \figref{fig:compare_falloff} we highlight a few visible artifacts on another character demonstrating apparent volume loss in the muzzle and elbow, as well as extraneous deformation of the cape. In
\figref{fig:knight} we \rev{s}how that solely applying the sparse control to the falloff-only setup by employing \myeqref{eq:voronoi_field_fn} but still keeping other parameters fixed, we are able to recover reasonable quality weights. Here, the sparsity control allows the model to selectively prune contributions between regions without relying solely on $r_j$, which affects falloff equally in all directions. In this case, the full SkinCells method converges 3.5x faster to a result shown in \figref{fig:knight}c maintaining good quality deformation. 
}

\begin{figure}
    \centering
    \includegraphics[width=\linewidth, trim={0 0 5 0}, clip]{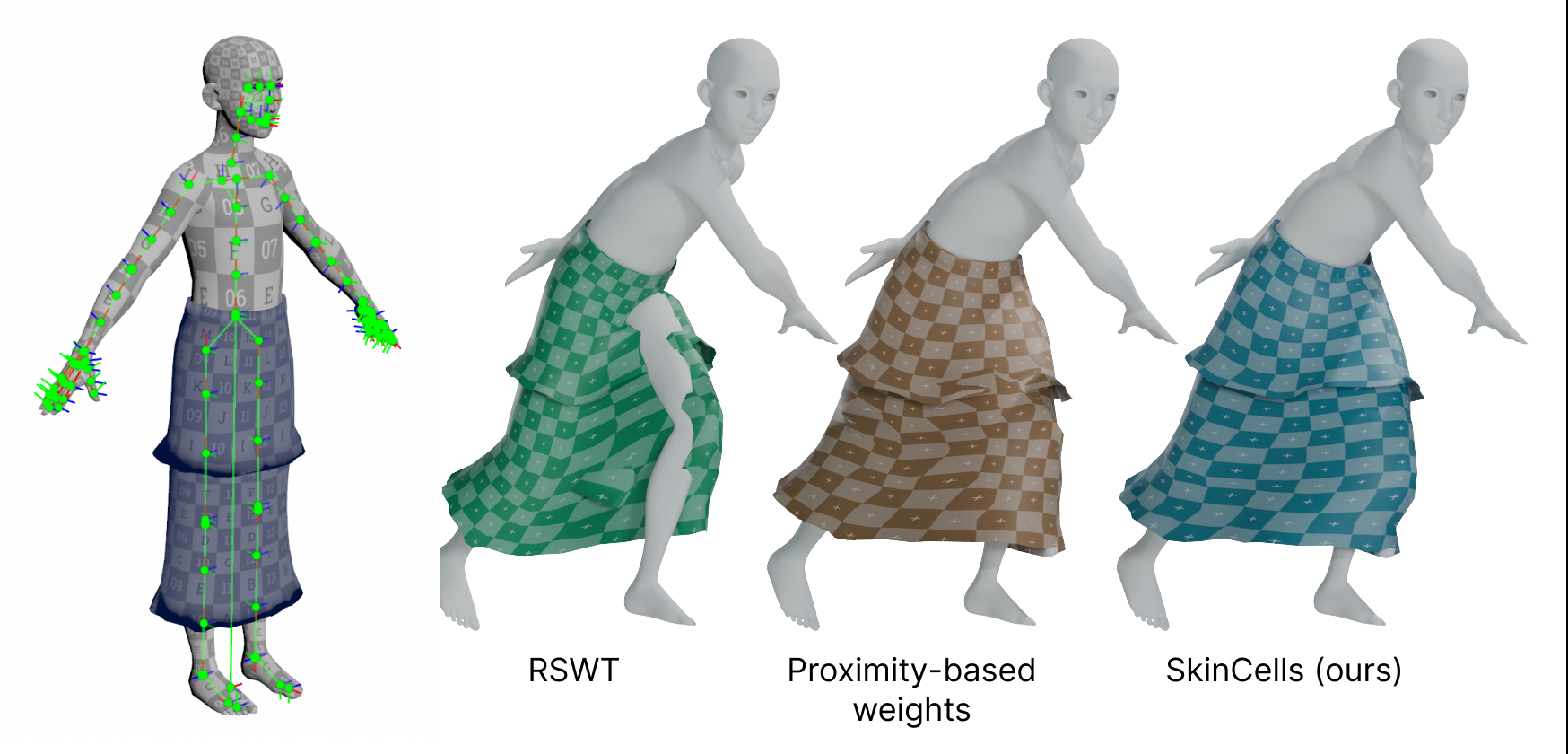}
    \caption{We animate a skirt garment on a character with a dedicated skirt rig attached to the pelvis (left). Our optimization-based method significantly reduces excessive stretching, outperforming a standard proximity-based technique. In addition, we compare to RSWT~\cite{abdrashitov2023} to show how weights transferred from the body fail to deform loose garments causing stretching and interpenetration. }
    \label{fig:cloth}
\end{figure}

\paragraph{Cloth skinning} High-quality skinning weights are crucial for garments like loose clothing and long dresses, where disturbing artifacts are easily identified. Long dresses and skirts often suffer from abrupt splitting along the middle due to ambiguous leg mapping or frequent body intersections. To address this, we add a collision penalty term to the total loss (\myeqref{eq:l}) \cite{stuyck2024quaffure}, reducing body-clothing intersections.
Our method produces noticeably less garment stretching compared to a proximity-based method as shown in Fig.~\ref{fig:cloth}.
The proximity-based approach is implemented by identifying the $l$-nearest bones for each vertex, then assigning weights to these bones based on a distance $d$ with a falloff of $1/d^{3.5}$, and finally normalizing the assigned weights to ensure they sum up to 1.
We also compare against robust skinning weight transfer (RSWT)~\cite{abdrashitov2023} where pre-determined weights on the body are used to drive the clothing. Compared to RSWT, our method is able to better resolve intersections with the body, and allow the cloth to follow a dedicated skirt rig.

\begin{figure}
    \centering
    \includegraphics[width=\linewidth]{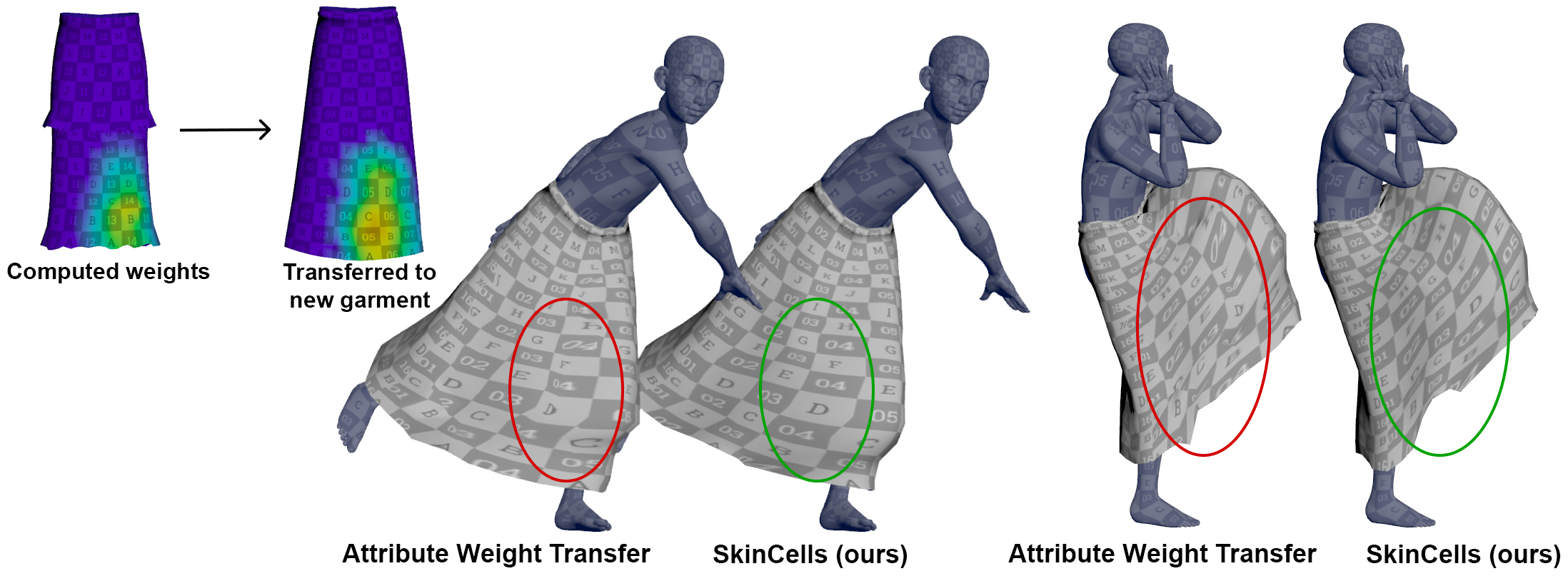}
    \caption{We demonstrate how our method can aid in transferring weights between similar meshes. Here we transfer a set of precomputed weights from one skirt to another using Houdini (top left). We demonstrate that transferred weights  can produce spurious stretches in the cloth, whereas applying our precomputed weight field to a similar garment can produce a smoother result.}
    \label{fig:attrib_transfer}
\end{figure}

\paragraph{Levels-of-detail and weight transfer}
Our method optimizes a spatially varying weight field for a garment mesh, which can then be applied to generate skinning weights for a higher-resolution, detailed mesh with small features like buttons or zippers. In \figref{fig:attrib_transfer}, we demonstrate transferring weights from one skirt to a similar skirt. This example shows that standard proximity-based attribute transfer techniques fall short when aiming for high quality deformation. Using the weight field produced with our method, one can simply evaluate it at the locations of the vertices of the new mesh to produce a smoother result.
Another application of this feature is automatically skinning meshes with varying levels-of-detail, where a single optimization is sufficient to provide weights for all levels. \rev{In \figref{fig:lod}, we show the optimization result using the mesh labeled ``Original'' with 9k vertices applied to a coarser mesh (left) and finer meshes (right), showing no visible skinning artifacts.}





\section{Discussion}

In early experiments, we used a multi-layer perceptron (MLP) to generate the weight field similar to existing popular approaches. However, we found that MLPs could not reproduce the sparsity needed to avoid artifacts after clamping to a specified number of bones per vertex, even if minimizing $\Loss_{\text{sp}}$. This originally inspired the development of SkinCells, which is substantially more compact than a comparable MLP and uses intuitive parameters that allow for effective initialization, thus reducing optimization time.

Another simple alternative to SkinCells is optimizing weights on the vertices of meshes directly, however we found that such methods \rev{are} less flexible and tend to converge to nearby local minima, thus struggling to make improvements on poor initializations.


\section{Limitations and Future Work}

\paragraph{Canonical pose} The Voronoi structure we use binds contiguous sections of space to the same bones, meaning that two points closer to one set of Voronoi sites than another will bind to the same bone. This being defined in 3D canonical space (as opposed to perhaps texture space) allows the algorithm to naturally bind surface elements that are close together to a similar set of bones. While generally this is desired, in some cases it can cause difficulties in distinguishing distinct parts of the body that are close together in canonical space. This is a well known limitation for any method that defines weight fields in canonical 3D space \cite{lin2022fite, li2024animatablegaussians, saito2021, bhatnagar2020loopreg}, and the recommended solution is to reconfigure the canonical representation to have large gaps between limbs and digits (e.g. using a T-pose with legs spread forming an ``A''). While this can be more challenging for fantastical characters and garments, we expect that an erosion algorithm to deform the geometry until all parts are well separated could help.

\paragraph{Self-contact} Our method does not account for self-contact when optimizing the skinning weights for garments. While self-contact can substantially improve the appearance of clothing in simulation, skinning is typically constrained to very few degrees of freedom that is usually insufficient to resolve self-intersections.

\paragraph{Artist control} While we designed our method to be fully automatic to enable automatic asset generation at scale, we acknowledge the benefit of optionally providing user-control to the generated results. Currently, one can manipulate the result by modifying the parameters associated with the optimization procedure, in the future we plan to leverage artist-painted weight maps to guide the optimization, offering more refined control.

\paragraph{3D representations} The optimized skinning weight fields can be used to animate other 3D representations such as point clouds or volumetric primitives like Gaussian Splatting \cite{kerbl3Dgaussians}, but the optimization process relies on the mesh connectivity to evaluate the smoothness loss. We \rev{plan} to explore alternatives to this loss to make our method even more general and support novel representations such as Gaussian avatars \cite{li2024animatablegaussians} \rev{in future work}.

\section{Conclusion}

We introduce a fully automatic skinning weight optimization method that requires only the minimal amount of data necessary for skinning: a skeleton rig and character mesh. This approach eliminates the need for additional user input, enabling the efficient generation of optimal skinning weights for a vast number of virtual characters. Due to its fully automatic nature, our method is particularly well-suited to complement recent advances in automatic mesh generation.
We propose a novel smoothness objective function that is minimized under automatically sampled poses, resulting in high-quality skinning weights. Our method is capable of enforcing a sparse set of bones influencing each vertex and our spatial nature for the weight definition allows for direct application to all resolutions in an LoD system. Through comprehensive evaluations, we demonstrate that our approach provides a robust solution to creating high quality skinning weights.

\section*{Acknowledgements}

We would like to thank Petr Kadlecek for his assistance with generating and formatting assets, and Edith Tretschk for her valuable feedback on neural methods as well as her help with implementation and code optimization. We are grateful to Nicky He for sourcing motions and poses used in the early stages of training, and to Yu~Ju (Edwin) Chen for insightful discussions on numerical methods and for narrating the supplemental video. Special thanks go to Rob Stratton for creating the manual skinning baselines for the initial skirt assets, Nicholas Burkard for helping with technical direction, and Gabriele Pellegrini for providing both art feedback and character skinning baselines. We also thank Anthony Yon for creating the skinning for the washing machine example, and Carlos Aliaga for his help in setting up rendered scenes.

\bibliographystyle{ACM-Reference-Format}
\bibliography{ref}

\end{document}